\documentstyle[11pt,newpasp,twoside,epsf]{article}

\begin{document}
\title{Pulsar Astrometry with the VLBA}
\author{Walter Brisken}
\affil{National Radio Astronomy Observatory, PO Box O, Socorro, NM 87801}

\begin{abstract}
Many features of the Very Long Baseline Array (VLBA) contrive to make it
the best telescope for pulsar astrometry.  The measured proper motions and
parallaxes allow distances and transverse velocities to be determined.  These
in turn provide clues to questions spanning nuclear astrophysics on
scales of $10^{-23}$\,m to the distribution of gas in the Galaxy on scales
of $10^{20}$\,m.  Three pulsars are discussed in this paper. 
Among pulsars, B0950+08 has the most accurate VLBI-determined parallax.
B1133+16 has a very high transverse velocity; its radial velocity is discussed.
B0656+14 is a thermally detected neutron star.  Determination of its distance
has allowed its radius to be measured and its association with the 
Monogem ring supernova remnant (SNR) to be established, 
allowing a long-standing question in cosmic ray astrophysics to be addressed.
\end{abstract}

\section{Introduction}
Pulsars
typically have velocities in the 100 to 1000\,km\,s$^{-1}$ range and 
preferentially populate the galactic plane.  Thus sub-milliarcsecond
astrometry is appropriate to determine distances and velocities of nearby
($\la 3$\,kpc) pulsars.  Most pulsar observations are forced to low
radio frequencies ($< 2$\,GHz)
in order to take advantage of their steeply inverted 
spectra ($S \propto \nu^\alpha$ with $\alpha$ typically in the range
-2.5 to -1.5).

The VLBA is uniquely suited for astrometric pulsar observations.  Parallax
measurements require observations roughly
every six months (but preferably every 
three months), making a full-time dedicated array such as the VLBA
preferable.  
The wide field of view of the identical 25\,m antennas allows one to find
and make use of
in-beam calibrators with a success rate of about 70\% at $\lambda=20$\,cm.  
These nearby 
calibrators make high-precision astrometry much simpler.  The VLBA correlator
has a pulsar gate allowing the signal-to-noise of pulsar observations to 
be increased by a factor of between about three and five, 
depending on the narrowness of
the pulsar's pulse profile, by disabling correlation during the off-pulse
portion of the pulsar's period.  

\section{VLBI astrometry}
Astrometry is the science of precise localization of astromical objects.  
Several 
measurements of an object over the span of a year or more can yield the
proper motion and annual parallax of the object.  VLBI is well suited 
for precision position measurements because of the high resolution
attainable with content-sized baselines.  
The phases, $\phi$, of the measured
visibilities are directly related to the location of the object:
\begin{equation}
\phi(\nu) = \frac{\nu}{c}(u\,l + v\,m) + \epsilon(l, m, t, \nu).
\end{equation}
Here $c$ is the speed of light, $\nu$ is the observing frequency, 
$(u, v)$ is the projected baseline vector, and $(l, m)$ is the location
of the object being observed relative to the correlator model phase
center.  Added to the phase is an error term, $\epsilon$, that varies 
with direction, time ($t$), and frequency. 
Phase-referencing is almost always used in VLBI astrometry. 
This form of relative astrometry observes a calibrator source with 
an accurately known position in a {\em nodding} cycle with
the target (pulsar).  At 20\,cm with the VLBA, this cycle typically consists
of about 90\,s on the calibrator followed by about 120\,s on the 
target.  This cycle is repeated dozens of times per observation.  
Since the position of the calibrator source is well known and 
placed at the phase center during correlation, its visibility
phases can be used to estimate the phase error:
\begin{equation}
l_{\mathrm{cal}} = m_{\mathrm{cal}} = 0 \longrightarrow
\epsilon_{\mathrm{cal}} = \phi_{\mathrm{cal}}.
\end{equation} 
Since the calibrator was chosen to be nearby the target, and the observations
of the target and calibrator are interleaved on a timescale short compared
to that of the evolution of $\epsilon$, the remaining phase error that 
remains after removing the interpolated calibrator phase from the target
phase is much reduced:
\begin{equation}
\Delta \epsilon = \epsilon - \epsilon_{\mathrm{cal}}
\ll \epsilon.
\end{equation}
While much reduced from the initial phase error, $\Delta \epsilon$
can remain substantial.  This remaining error term is usually dominated by
tropospheric and ionospheric gradients.  The troposphere introduces
a frequency independent delay, which corresponds to a phase error 
that is proportional to frequency.  The ionosphere has a dispersive
delay with a phase inversely proportional to frequency.  
Thus to first order, 
\begin{equation}\label{eqn:de}
\Delta \epsilon = \underbrace{A(l, m, t)\,\nu}_{\mathrm{Troposphere}} +
		  \underbrace{B(l, m, t)\,\nu^{-1}}_{\mathrm{Ionosphere}}.
\end{equation}
Two methods have been used in pulsar astrometry
to improve phase-referenced astrometry.  The first aims at minimizing
the uncalibratable errors and the second aims at modeling and removing dominant
errors.

\subsection{In-beam calibration (Chatterjee et al., 2001)}

Since both $A$ and $B$ increase with increased target-calibrator separation,
it is advantageous to find a calibrator as near the target as possible.
If the target and calibrator can both be placed within the primary beam
of all of the antennas in the array ($\sim 25\arcmin$ at 20\,cm for the VLBA),
then both sources may be simultaneously imaged.  The simultaneity
also improves the phase-referencing.  Two correlator passes are
needed in almost all cases.  

\subsection{Ionosphere removal (Brisken et al., 2000)}

For $\lambda \ga 6$\,cm the ionosphere dominates the phase error.  At 
$\lambda \sim 20$\,cm, the ionosphere is responsible for about 90\% of the
phase error.  Because the
ionospheric phase errors have a differant frequency dependence than 
the tropospheric phase errors and the geometry (i.e., the pulsar's
location), observations made over a substantial fractional bandwidth allow
the ionosphere strength above each antenna, $B$, to be deduced.  
The frequency agility of the VLBA allows 8 independently tunable
spectral windows to be placed within a 500\,MHz portion of a band,
providing enough leverage to measure and remove the ionospheric
component of phase error.
This method works well for target-calibrator separations up to about
3.5\deg, as long as both sources are strong enough and do not exhibit
structure that changes significantly across the observed band.

\section{Astrometry of PSR B0950+08 (Brisken et al., 2002)}

\begin{figure}
\plottwo{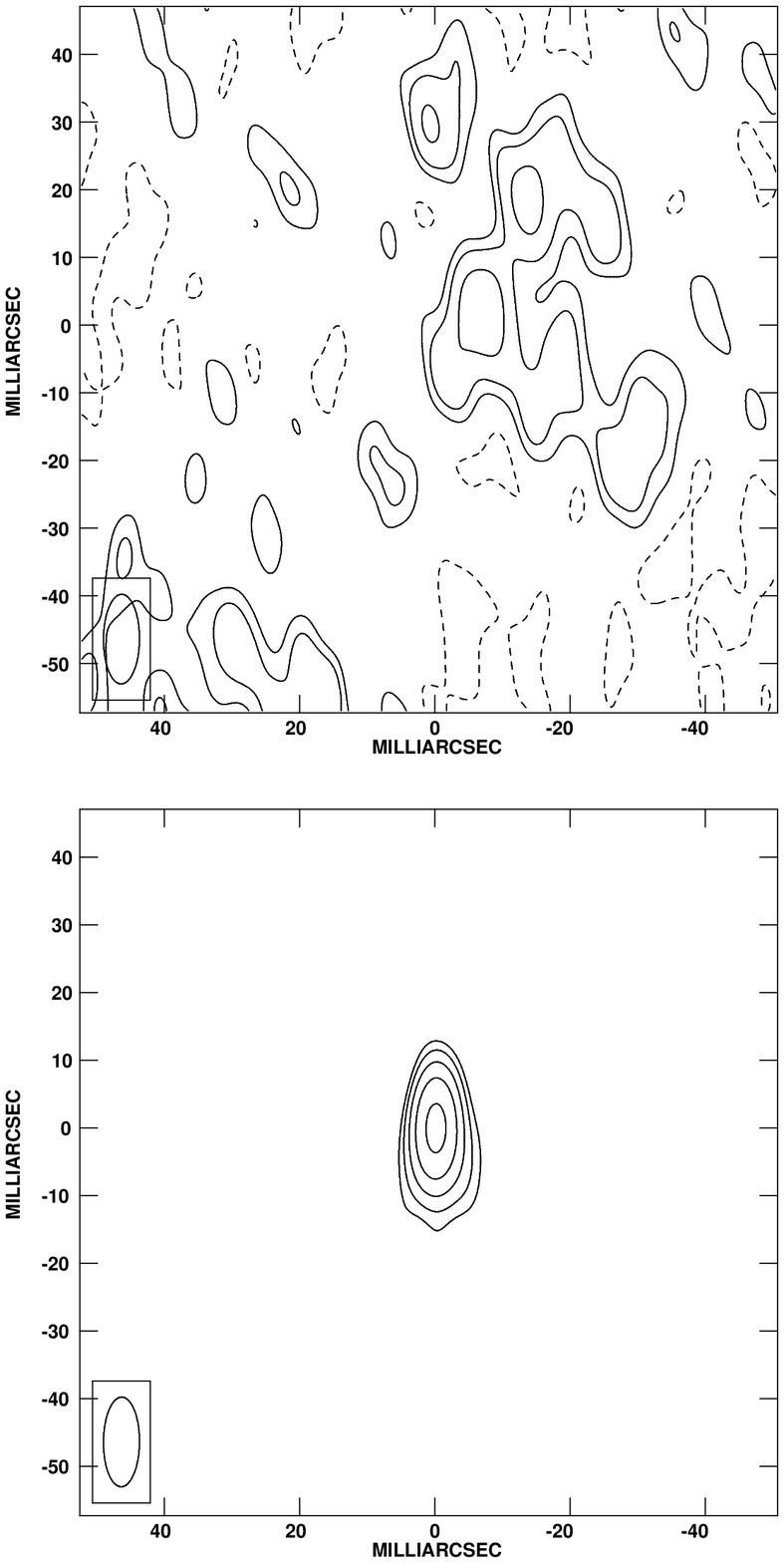}{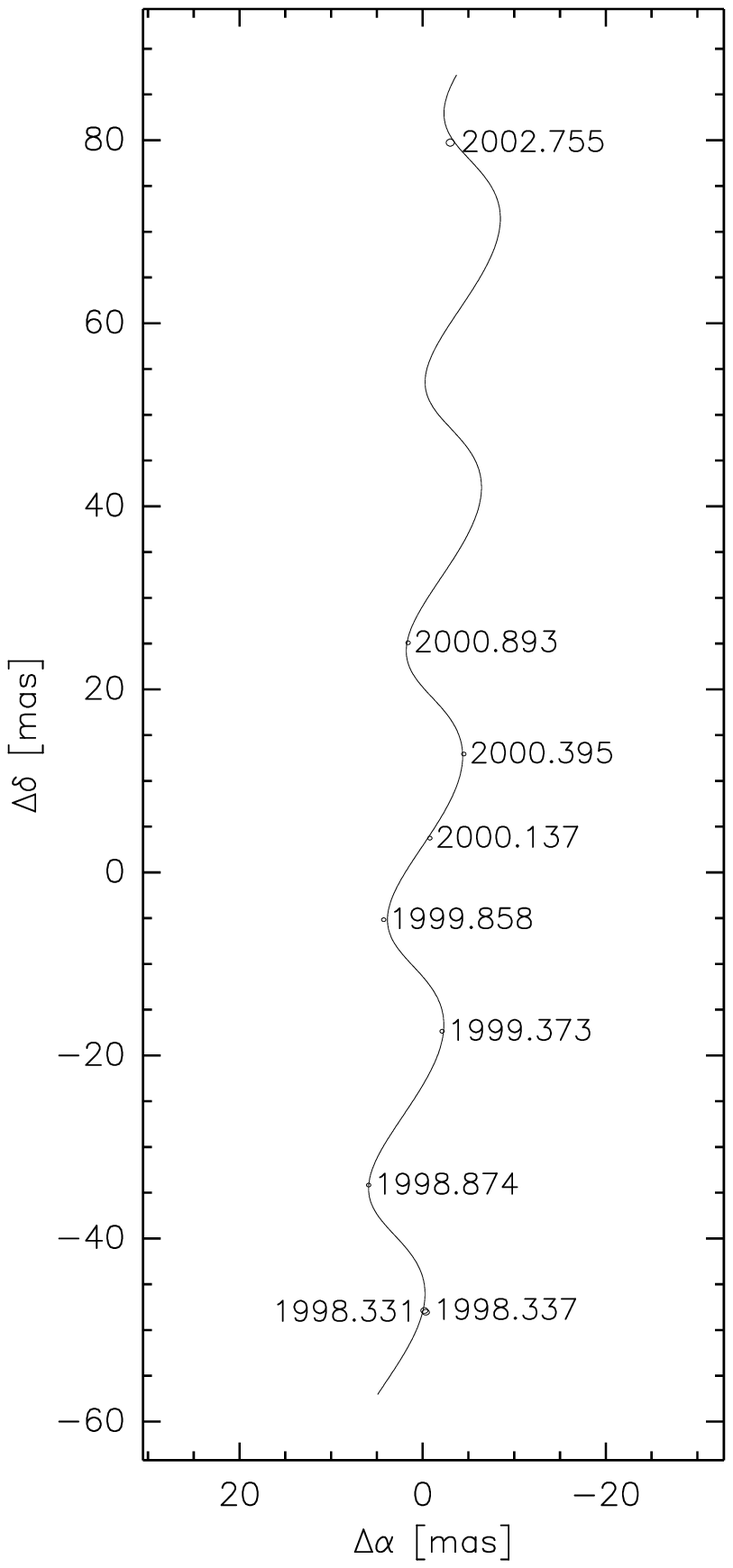}
\caption{
\label{fig:b0950}
Ionospheric removal applied to B0950+08.  On the left is an
example of an image of B0950+08 before (above) and after 
(below) removal of the 
ionospheric phase errors.  The contours are the same in each figure, increasing
by factors of 2 from the lowest contour at 5\,mJy\,beam$^{-1}$.
On the right are the measured positions of B0950+08 at 8 different
epochs.  The best fit parallax and proper motion model is shown as the solid
line.  Note that the sizes of the ellipses are representative of the 
measurement uncertainties.}
\end{figure}

In 1998 and 1999, VLBA observations of PSR B0950+08 were used to develop
the ionospheric removal technique.  A demonstration of the importance of
improving phase-referencing 
is shown in Figure~\ref{fig:b0950}.  After three successful epochs of 
observation, a parallax for B0950+08 was determined and a project to 
measure parallaxes of ten pulsars began.  
The proper motion and parallax measurements
of B0950+08 improved considerably 
with the addition of four more epochs.  Eight of the other 
pulsars each had four or five observations over the course of one year,
resulting in parallaxes and proper motions.  The
tenth pulsar was never detected.  

An eighth observation of B0950+08 occured in October 2002 as part of NRAO's
Mark~5 development.  The Mark~5 disc-based VLBI recording system is
being developed by Haystack Observatory as a replacement for the aging
and bandwidth-limiting tape recorder systems currently deployed at almost
all VLBI antennas.  Mark~5 promises increased recording bandwidth,
higher reliability and less expensive media.  Although this additional 
observation was made using only Pie Town, Hancock and St. Croix antennas
and did not employ the pulsar gate, it did further constrain the proper
motion of B0950+08.  A final fit to all eight epochs yielded a (J2000)
proper motion of $\mu_\alpha \cos \delta=-2.06\pm0.07$\,mas\,yr$^{-1}$,
$\mu_\delta=29.37\pm0.05$\,mas\,yr$^{-1}$, and a parallax of 
$\pi=3.81\pm0.07$\,mas (all quoted uncertainties in this paper are 
68\% confidence intervals).
A plot of the motion of B0950+08 and its proper motion and parallax fit
is shown in Figure~\ref{fig:b0950}.

\section{The radial velocity of B1133+16}

B1133+16 is a bright, fast moving pulsar with characteristic age 
$\tau_{\mathrm{char}} = 5$\,Myr.
Its distance, $D = 357^{+22}_{-19}$\,pc, and transverse velocity, 
$v_{\perp} = 636\pm40$\,km\,s$^{-1}$ were measured with the VLBA using the
ionosphere removal technique (Brisken et al., 2002).

\begin{figure}
\plotone{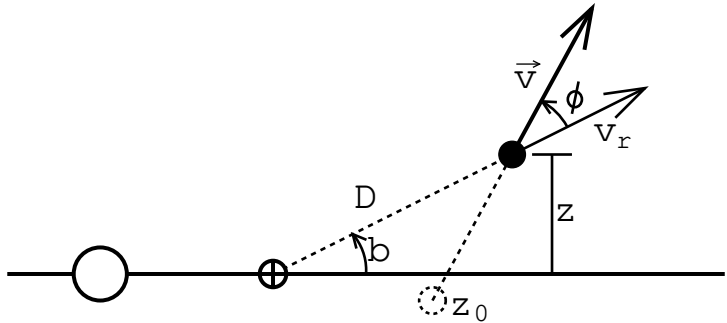}
\caption{
\label{fig:b1133}
The geometry used to calculate radial velocities.  The horizontal
line represents the plane of the galaxy; the large circle is the
Galactic center.  Earth is the circle with the `+'
in the middle.  The pulsar is born at the location of the dashed circle and
is currently at the filled circle.}
\end{figure}

The radial velocities of pulsars are unmeasurable, although in a few cases
they can be inferred.  B1133+16 has a high galactic latitude, $b = 69.2$\deg.
If the true age $\tau$ and the birth height $z_0$ are known,
then the radial velocity $v_r$ can be inferred from its current height
above the galactic plane, $z = D \sin b$:
\begin{equation}
v_r = \frac{D \sin b - D \mu_b \tau \cos b - z_0}{\tau \sin b}.
\end{equation}
See Figure~\ref{fig:b1133} for a diagram of the geometry.
The birth height is not known, so a sensible birth height distribution
must be assumed.  Here $P(z_0) \propto e^{-|z_0/a|}$ with scale height
$a = 150$\,pc is used.  Likewise the true age is not known with certainty, so
an age distribution based on the timing age is assumed as well.
The probability distribution for $v_r$ is then given by
\begin{eqnarray}
P(v_r) & = & \int dz_0 \int dD \int d\tau \int d\mu_b \ 
\delta\left(v_r - \frac{D \sin b - D\mu_b\tau\cos b - z_0}{\tau \sin b}\right) 
\nonumber \\
& & \times \,P(z_0)\,P(D)\,P(\tau)\,P(\mu_b).
\end{eqnarray}
For many pulsars, this is quite non-constraining, but in the case of B1133+16,
the result is significant.  The inferred radial velocity is
$v_r = -45^{+60}_{-40}$\,km\,s$^{-1}$.  That this is quite small compared
to the transverse velocity means that this pulsar must be moving very nearly
in the plane of the sky, $\phi = 99 \pm 9$\deg.

\section{PSR B0656+14}

PSR B0656+14 is an adolescent pulsar with timing-derived age of 110\,kyr.
A five-epoch observing campaign was started with the goal of determining
its distance to 10\%.  With a distance measurement, its thermal
spectrum would reveal its radius.  

\subsection{Distance and Radius (Brisken et al., 2003)}

Using a nearby bright in-beam calibrator,
a (J2000) proper motion of 
$\mu_\alpha \cos \delta=44.07\pm0.64$\,mas\,yr$^{-1}$,
$\mu_\delta=-2.40\pm0.29$\,mas\,yr$^{-1}$, and a parallax of
$\pi=3.47\pm0.36$\,mas were fit to the five epochs of data.  The
distance derived from this measurement, $288^{+33}_{-27}$\,pc, is
less than half the distance estimated from its dispersion measure.

Several pulsars and radio-quiet neutron stars are known to emit blackbody
radiation.  A distance measurement combined with the spectral energy
distribution can thus yield a blackbody radius.  Observations in the optical,
UV and x-ray are fit well by a three component model consisting of
a soft blackbody representing the neutron star surface 
with a temperature of $\sim 8.2$\,K 
($8.4\pm0.3\times10^5$\,K, Koptsevich et al., 2001; 
 $8.0\pm0.3\times10^5$\,K, Pavlov et al., 2002),
a hard blackbody representing the polar cap regions with a temperature of
$1.6\pm0.3\times10^6$\,K (Pavlov et al., 2002), and a power law spectrum,
presumably from the magnetosphere.  If it is assumed that a pure blackbody
accurately describes the emission from the surface of the neutron star, then
the VLBA distance of 288\,pc would imply an unrealistically small 
observable radius, $R_\infty \equiv R(1-2GM/Rc^2)$, of between 6.9 and 8.5\,km.

A blackbody spectrum becomes distorted in the presence of an atmosphere.
Several neutron star atmosphere 
models have been considered, but most of these
produce radii that are either implausibly small or large.  The most
realistic radius results from application of a magnetic hydrogen
atmosphere (e.g., Shibanov et al., 1993).  A range of radii are
supported by the various magnetic hydrogen atmosphere models present
in the literature, but all lie in the range $\sim 13$ to $\sim 20$\,km.
This range of radii does not usefully constrain the equation of state of
matter at nuclear density, but additional phase-resolved optical 
observations and an improved distance through additional VLBA observations
will likely shrink this range considerably.


\subsection{Connection with the Monogem Ring (Thorsett et al., 2003)}

The Monogem ring is a $\sim 25\deg$ diameter soft x-ray shell.  Although this 
source has been speculated to be a SNR, positive identification has been
difficult as the source is not well seen in other wavebands, especially the
radio.
If this source is assumed to be a SNR, then
modelling suggests a distance
of about 300\,pc and an age of about 86\,kyr.  Earlier attempts to 
associate B0656+14 with this SNR have been dismissed because of the
pulsar's 760\,pc dispersion measure distance.  The VLBA distance measurement
of B0656+14 agrees 
well with the SNR distance, and the ages are consistent with one another.
Thus it is claimed that the same supernova event created the Monogem ring
and B0656+14.

\subsection{The cosmic ray spectrum `knee' (Thorsett et al., 2003)}

Above $10^{10}$\,eV, the cosmic ray spectrum has only two distinct features:
a {\em sharp} steepening at $\sim 3 \times 10^{15}$\,eV called the `knee',
and a flattening around $3 \times 10^{18}$\,eV called the `ankle' 
(see Wefel 2003 for a review).  Although many possible explanations have 
been given, the sharpness of the knee has been difficult to explain.
Erlykin \& Wolfendale (1997) noted that a single dominant nearby source,
rather than a superposition of many sources with varying properties or
instrumental/propagation effects, would most easily explain the
sharpness.  They suggested that the knee could be a cosmic ray excess 
due to 
accellerated oxygen and iron nuclei from a 90 to 100\,kyr old,
300 to 350\,pc distant SNR.
The Monogem ring, now quite firmly established as a SNR of nearly the
right age and distance, is proposed as the likely source.

\section{Conclusion}

Pulsar astrometry has made great advances with the VLBA.  Eleven of the
thirteen VLBI pulsar parallax measurements have occured in the last 3 years 
using the VLBA.  Techniques have been demonstrated on ordinary pulsars,
such as B0950+08, pulsars with special interests have been targeted,
such as B0656+14, and ongoing is a VLBA large project aiming to increase
by a factor of two or three 
the sample of pulsars with parallax derived distances.  Accurate
distances and velocities of a sizable number of pulsars will allow the
velocity distribution of pulsars to be probed with greater precision than
currently possible and will improve Galactic electron density models, which
will allow better estimates of distances to pulsars that do not have
measured parallaxes.  Accurate astrometry also promises to improve the
statistics on spin-velocity alignment.  Much increased bandwidth will likely
be available within the next decade, bringing many more interesting
pulsars within the sensitivity limits of the VLBA.

\end{document}